\documentclass[letterpaper, 10 pt, conference]{ieeeconf}  

\IEEEoverridecommandlockouts                              

\usepackage{graphics} 
\usepackage{epsfig} 
\usepackage{mathptmx} 
\usepackage{times} 
\usepackage{amsmath} 
\usepackage{amssymb}  
\usepackage{url}
\usepackage{multirow}
\usepackage{textcomp}

\title{\LARGE \bf
Development of a system to profile foot temperatures of the plantar and the periphery
}

\author{$^{1}$Madarasingha K.C.M., $^{1}$Perera W.N.D., $^{1}$Rathnayaka A.M.A.I., $^{1}$Savindu H.P., $^{2}$Jayasinghe S., \\
$^{3}$Kahaduwa K.T.D., $^{1}$De Silva A.C., \textit{Member}, IEEE  
\thanks{This work was supported by the National Science Foundation grant RPHS-2016-DTM02.}
\thanks{$^{1}$Department of Electronic and Telecommunication Engineering, Faculty of Engineering, University of Moratuwa, 10400 Katubedda, Sri Lanka.
        {\tt\small chamarakcm@gmail.com}}%
\thanks{$^{2}$Department of Clinical Medicine, Faculty of Medicine, University of Colombo, Sri Lanka.}%
\thanks{$^{3}$Professorial Surgery Unit, National Hospital, Sri Lanka.}%
}

\begin{document}

\maketitle
\thispagestyle{empty}
\pagestyle{empty}

\begin{abstract}
Foot temperature profiling is of utmost importance is mitigating the adverse effects due to foot complications especially due to diabetes. Contactless temperature monitoring methods could be used effectively in large scale for patient screening. Near-infrared thermography has proven to be convenient and accurate for temperature profiling. The objective of this study is to develop a diagnostic device using the said imaging technology to detect as well as progress monitoring of foot complications. The device we have developed is capable of scanning the foot plantar and the periphery and it is also accompanied by a semi-supervised thermal image analysis algorithm which is convenient to the clinician. Preliminary clinical testing conducted using 6 diabetic subjects out of which 2 had ulcers in either foot and 9 non-diabetic subjects 2 of which had wounds on the plantar. The system was able to detect the ulcerated areas and wounds with the algorithm developed specifically for thermal image analysis. 
\end{abstract}


\section{INTRODUCTION}

Foot temperature monitoring is a promising modality for early detection of foot complications such as ulceration and infection. The main causes of foot complications are two conditions  ‘Neuropathy’ (nerve damage) and ‘Ischaemia’ (problems with the supply of blood to the feet) associated with diabetes \cite{singh2005preventing}. Over the past three decades from 1980 to 2014, the worldwide population of diabetic patients has risen from 108 million to 422 million which is almost a fourfold increment which shows the significance of the disease as one of the most common non-communicable diseases in the world \cite{world2016global}. The World Health Organization (WHO) predicts diabetes to be the world’s seventh leading cause of death by 2030 \cite{mathers2006projections}.

All types of diabetes can result in complications in several parts of the body and also increase the risk of premature death. The complications of diabetes can be heart attack, stroke, blindness, kidney failure, lower limb amputations, nerve damage and fetal death during pregnancy \cite{world2016global}.

Early detection of ulcerations and inflammations is possible through foot temperature monitoring since inflammation which precedes ulceration and infection results in a temperature increase at the foot surface.
Since the temperature increase is only witnessed in the affected area of the foot, a difference is observed when the temperature of that location is compared with the same location on the contralateral foot. Through early detection of foot complications, required medical intervention can be made at an earlier stage to prevent further adverse consequences, morbidity, and mortality. Through research investigations, it has been validated that the most optimal cutoff skin temperature value for detection of diabetes-related foot complications as 2.2\textsuperscript{0}C \cite{van2014diagnostic}.
 
The objective of our study is to develop a device to monitor the foot temperature of the plantar and the periphery of the foot and along with its software tools to predict the formation of diabetes-related foot complications. The device is an electro-mechanical system inclusive of thermal cameras which follows an automated thermal image acquisition procedure to capture thermal images of the plantar and the periphery of the foot. The developed software tools will analyze the captured thermal images and pinpoint the locations of the feet where there is a high risk of ulceration.
 
\subsection{Existing Technologies and their Drawbacks}
Numerous research projects have been conducted focusing on measuring/monitoring the foot plantar temperature and to analyze their correlation towards early detection of foot complications. These research studies have used contact and contactless methods to image plantar temperature distributions.  

The study to find the optimal diagnostic cutoff skin temperature value for detection of diabetes-related foot complications used FLIR\textsuperscript{TM} thermal camera and determined the optimal diagnostic cutoff skin temperature value as 2.2\textsuperscript{0}C. However, in their discussion, few improvements are suggested such as detecting temperature differences between left and right feet at any spot on the foot (not only at selected spots), analyzing the size of detected hotspots etc. which have been achieved in the system developed by us [4]. The study to find the correlation between plantar foot temperature and diabetic neuropathy was a case study which used infrared thermal imaging technique to image the temperature distribution \cite{bagavathiappan2010correlation}.

There are studies which used contact sensors to measure the foot temperature of specific areas \cite{murillo2014foot}. However, contact methods only provide temperatures of several selected point but not the overall foot. Additionally, images are less accurate, less sensitive, slower and possess hygienic issues when applied to larger patient samples. Therefore, contactless imaging modalities are preferred when taking temperature measurements. Moreover, there are several commercially available products which measure point temperatures of the skin. Two such devices are TempTouch \cite{TempTouc71:online} and MT4 Diabetik \cite{NonConta71:online}. However, using these devices we are unable to construct a comprehensive image of foot temperature distribution since they only measure point temperatures.
 
\subsection{Proposed Solution}
The proposed solution is to develop an automated electro-mechanical system to acquire the thermal images of the plantar and the periphery of the foot using near-infrared thermal (NIR) cameras and to develop software tools to carry out the thermal image analysis and identify the areas with a higher risk of diabetes-related foot
complications.

It has also been noted that there is a considerable risk of ulceration in the periphery of the foot in addition to the foot plantar region \cite{LegandFo99:online}. Therefore, it is important to include the study of the periphery of the foot which is a novel idea to the existing literature on devices to measure foot temperature. The proposed device will generate a thermal image of the foot plantar and its periphery which is an advancement to the existing foot temperature measuring systems.

\section{METHODOLOGY}
The temperature scanning system consist of two main components; The image acquisition system (hardware system) and the image analysis system (software system)

\subsection{Image Acquisition System}
%
%

\begin{figure}[thpb]
      \centering
      \includegraphics[scale=0.5]{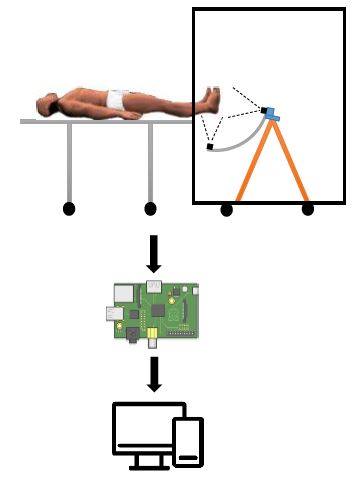}
      \caption{Overall system: Image capturing device, Microcontroller unit and PC}
      \label{fig:device}
   \end{figure}

The image acquisition system was developed to include the functionality of acquiring both the foot plantar and its  periphery images. Since NIR thermography was the preferred method for sensing the temperature, two FLIR\textsuperscript{TM} Lepton 3.0 Long-wave Infra-red cameras with radiometry were used. The spectral range of the thermal camera is 8-14$\mu$m and this range includes the normal human IR (Infra-red) radiation wavelength which is about 12$\mu$m \cite{HKObservatory}.  The camera had a resolution of 160$\times$120 pixels with a thermal sensitivity of 0.05K and a horizontal field of view of 56\textsuperscript{0} and a diagonal field-of-view of 71\textsuperscript{0}. The distance at which the camera gave the optimum image was calculated by taking the field-of-view angles into account which resulted in a 38cm $\times$  28cm image. The firmware for image capturing and camera controlling was developed by the software development kit provided by FLIR\textsuperscript{TM} and the software interface description document \cite{146:online}. A Raspberry Pi 3 model B development board was used to control the cameras and to interface the data acquisition system with the PC. One of the cameras was fixed capturing the foot plantar, while the other was mounted on a rotating C-arm capturing images of the foot periphery at four angles. A stepper motor rotated the C-arm which was also controlled by the Raspberry Pi. The camera system was mounted in a 1.5m$\times$1.5m$\times$1.5m box which was covered by a black textile to prevent interference from heat sources outside. The overall system is shown in figure \ref{fig:device}. Image data was transferred to the PC via a TCP/IP connection.

\begin{figure}[thpb]
      \centering
      \includegraphics[scale=0.45]{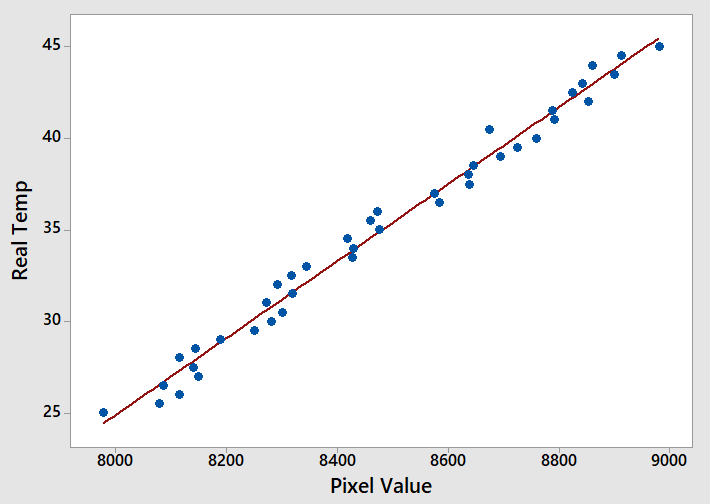}
      \caption{Calibration curve of the camera using least square fit}
      \label{fig:tempcalb}
   \end{figure}

The cameras were calibrated using the immersion thermostat 'Digit-Cool' (J.P. Selecta, Barcelona, Spain)   with a water bath. The accuracy of the immersion thermostat is 0.01\textsuperscript{0}C, which is higher than that of the thermal camera. The images of the water surface were captured at 0.5\textsuperscript{0}C intervals from 25\textsuperscript{0}C to 45\textsuperscript{0}C. The mean pixel values at each temperature were plotted against the actual temperature (see Fig. \ref{fig:tempcalb}) to obtain the calibration curve fitted using the least square approach. The non-linearity error of the camera sensor was found to be 2.96\% according to (\ref{eq1}).

\begin{equation}
Non-linearity (\%) = \frac{Maximum \ deviation \ in \ input}{Maximum \ full \ scale \ input}
 \label{eq1}
\end{equation}

\subsection{Image Analysis System}
\begin{figure}[thpb]
      \centering
      \includegraphics[scale=0.7]{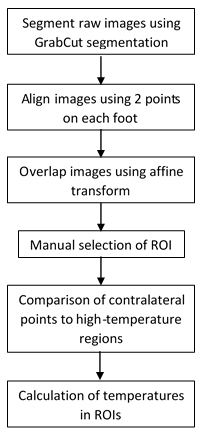}
      \caption{Flow diagram of data analysis}
      \label{fig:algo}
   \end{figure}

A custom developed software to analyze images was a key requirement from relevant clinicians. Therefore, as an integral part of the system, a semi-supervised image analysis software was developed with the objectives; 1) to detect areas with ulceration 2) to calculate the mean temperature (MT) of both the feet 3) to identify areas which are more prone to ulceration (also defined as regions of interest (ROI) e.g. toe, metatarsal and heel). The algorithm used for the analysis is shown in Fig. \ref{fig:algo} in the form of a flow diagram.
%
%
%
%

\subsection{GrabCut Image Segmentation}
GrabCut is an image segmentation method based on graph cuts which separates the background from the foreground with a minimum input from the user \cite{rother2004grabcut}. For this application, GrabCut was preferred over other segmentation methods because the user can intervene to obtain the correct segmentation, whereas fully automatic segmentation has a high chance of failing due to irregularities in the image boundary.

\subsection{Affine Transformation}
\begin{figure}[thpb]
      \centering
      \includegraphics[scale=0.75]{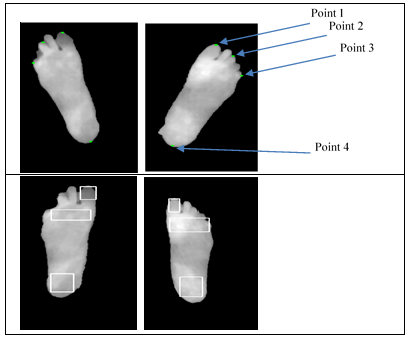}
      \caption{Selection of points for affine transform - Points 2 and 4 are considered for vertical alignment (top) selected ROIs after aligning two feet (bottom)}
%
%
      \label{fig:affineROI}
   \end{figure}

Since most of the patients tend to keep their feet at different orientations at the time of scanning, it was required to use affine transform to align the feet and overlap so that contralateral points of the feet can be compared. The user was allowed to select four points similar to what is shown in Fig. \ref{fig:affineROI} Top as the input to calculate the affine transform which is used for the following two tasks:

\begin{itemize}
\item Two points out of the four to aligning the feet vertically. This makes it convenient to define the regions of interest in which temperatures need to be assessed.

\item Three points out of the four to calculate the homography between the two images and overlap so that contralateral points can be compared to detect the high-temperature areas automatically. 

\end{itemize}

\subsection{Neighborhood Comparison}
However, considering only the difference is not sufficient since a region which has a lower temperature compared to the corresponding region on the other foot which is in normal temperature, may be classified as a high-temperature area, which implies that it is a possible ulcerated area. To avoid such misclassification, a method defined as `neighborhood comparison' was used as described below.

When two corresponding regions on two feet depict a temperature difference of 2.2\textsuperscript{0}C, the mean temperature (MT) of a larger region around the higher temperature region is calculated and compared with the original temperature.

%
%
\begin{itemize}
\item If the MT of the original region is similar to that of extended region, then the temperature difference is deduced to be caused by a colder region in the other foot.
\item If the MT of original region is greater than that of extended region, then the temperature difference is deduced to be caused by a possible ulcerated area.
\end{itemize}

\subsection{Experimental Setup}
Since the initial study was focused on verifying the device and its sensitivity to temperature changes resulting from wounds, the following groups were selected. 

\begin{itemize}
\item Diabetic patients with ulcers on a single foot and diabetic patients without ulcers on both feet
\item Non-diabetic patients without wounds in either foot and non-diabetic patients with wounds resulting from external injuries 
\end{itemize}

Subjects with a single ulcerated/wounded foot was selected since the initial algorithm compares the contralateral points of the feet of an individual when making a judgement. A total of 15 individuals including 6 diabetic subjects and 9 non-diabetic subjects participated in the preliminary test. MT of the entire foot, toe, metatarsal and the heel were measured using thermal images. Two of the diabetic subjects had foot ulcers and calluses. Two subjects of the non-diabetic sample had wounds on one of the foot plantar. 

\section{RESULTS}
Using the aforementioned data analysis procedure, areas of foot ulcers were detected  based on the temperature difference of 2.2\textsuperscript{0}C between contralateral points of the two feet.

Table \ref{table_MT_SD} shows the MT of the overall foot and the ROIs of diabetic subjects with and without ulcers.
According to the results, the difference of contralateral points in ulcerated subjects is significantly higher than diabetic subjects without ulcers despite the difference not being equal to 2.2\textsuperscript{0}C. Diabetic subjects with ulcers show MT values less than that of non-ulcerated subjects possibly due to the prevalence of cold feet which had already been clinically diagnosed.  

\begin{table}[h]
\caption{Mean Temperatures of Diabetic Subjects}
\label{table_MT_SD}
\begin{center}
\begin{tabular}{|p{1.4cm}|p{1cm}|p{0.9cm}|p{0.5cm}|p{0.7cm}|p{0.7cm}|p{0.5cm}|}\hline
\multirow{2}{*}{MT} & \multicolumn{3}{c}{Ulcerated Subjects} \vline & \multicolumn{3}{c}{Non-ulcerated Subjects} \vline \\\cline{2-7}
 & {Ulcerated Foot} & {Healthy Foot} & {Diff.} & {Left Foot} & {Right Foot} & {Diff.} \\\hline

\textbf{Toe} & 33.35 & 32.54 & 0.81 & 35.14 & 35.40 & 0.26\\\hline
\textbf{Metatarsal} & 33.84 & 32.96 & 0.88 & 35.51 & 35.70 & 0.19\\\hline
\textbf{Heel} & 34.41 & 32.89 & 1.52 & 35.09 & 35.56 & 0.47\\\hline
\textbf{Overall} & 33.60 & 32.06 & 1.54 & 35.19 & 33.19 & 2.0\\\hline
\end{tabular}
\end{center}
\end{table}

\begin{table}[h]
\caption{Mean Temperatures of Non-diabetic Subjects}
\label{nonDiab}
\begin{center}
\begin{tabular}{|p{1.3cm}|p{0.95cm}|p{0.9cm}|p{0.4cm}|p{0.6cm}|p{0.6cm}|p{0.4cm}|}\hline
\multirow{2}{*}{MT} & \multicolumn{3}{c}{Subjects with Wounds} \vline & \multicolumn{3}{c}{Subjects Without Wounds} \vline \\\cline{2-7}
 & {Wounded Foot} & {Healthy Foot} & {Diff.} & {Left Foot} & {Right Foot} & {Diff.} \\\hline

\textbf{Toe} & 33.65 & 32.92 & 0.73 & 31.25 & 31.39 & 0.14\\\hline
\textbf{Metatarsal} & 34.39 & 32.72 & 0.67 & 31.26 & 31.30 & 0.04\\\hline
\textbf{Heel} & 34.20 & 32.73 & 1.47 & 30.87 & 31.10 & 0.23\\\hline
\textbf{Overall} & 31.78 & 31.32 & 0.46 & 31.25 & 28.91 & 2.34\\\hline
\end{tabular}
\end{center}
\end{table}

\begin{figure}[thpb]
      \centering
      \includegraphics[scale=0.7]{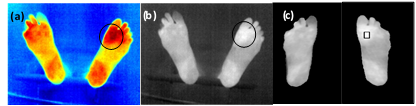}
      \caption{High temperature area in red color is highlighted by circle (a), High temperature area is highlighted in grayscale image (b), High temperature area is successfully identified by the semi-supervised analysis process and indicated using a rectangle (c).}
%
%
      \label{analysis_image}
   \end{figure}

Table \ref{nonDiab} shows the MT of the overall foot and the ROIs the non-diabetic subjects with wounds due to external injuries in one foot and without such wounds. The wounds due to external injuries were present in heel and/or metatarsal regions. Hence, the temperature values indicate that there is an aberrant increase in temperature in these regions. 
However, individual analysis of the images provide a more clear insight into the temperature difference and the regions. The software platform developed could detect the areas of ulceration and wounds. Fig. \ref{analysis_image} shows the results of the individual analysis performed using the system.

\begin{figure}[thpb]
%
%
      \centering
      \includegraphics[scale=0.8]{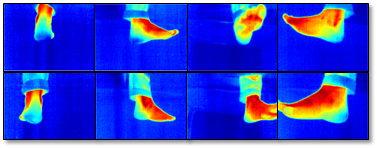}
      \caption{A-Thermal images of periphery of right leg, B-Thermal images periphery of left leg}
      \label{peri_images}
   \end{figure}

Fig. \ref{peri_images} shows the thermal images of the periphery captured at 90\textsuperscript{0} steps using the system developed. These images are included only to illustrate the capability of the developed system. An in-depth study on anomaly detection of the foot periphery needs to be done as a separate study.  

\section{DISCUSSION}
The results generated by this system depicts the temperature difference in a normal foot compared to an ulcerated/wounded foot. In fact, the preliminary objective of this study was to validate the ability of the device to detect abnormal areas of the foot caused by physiological anomalies. Hence, the ability of this device to distinguish between a normal foot and a wounded/ulcerated foot shows that the system is sensitive to such temperature changes. The semi-supervised analysis procedure used in the system is novel and convenient in the clinicians' perspective as there are no well documented procedures for thermal image analysis in current modalities according to the literature reviewed.

There is a wide range of applications for which this device can be used. Progress monitoring of ulcerated/wounded feet and monitoring the temperature changes in diabetic patients who are in the risk of ulceration are such key areas. 

Since this system currently employs comparison of two feet from the same patient, the ability of the system to analyze a patient with both feet wounded is questionable. Hence, the algorithm should be altered to tackle such situations. During the preliminary testing of the system, analysis of peripheral images was not performed.
Furthermore, the system could also make use of other important parameters such as foot pressure, which would enhance the accuracy of ulcer prediction.
%
%

The novelty of this study is the development of a device which can be used directly for clinical purposes and has the ability to capture thermal images of both the plantar and the periphery of the feet. Moreover, this system is capable of analyzing the plantar images and detect the possible areas vulnerable to foot ulcers.

The rotating C-shaped arm could accurately rotate 90\textsuperscript{0} angles focusing on main four sides of the periphery of the foot. Therefore together with images of the foot plantar, the temperature variation of the whole foot could be visualized using the GUI (Graphical User Interface) of the system. FLIR\textsuperscript{TM} Lepton 3.0 thermal camera, which is the latest version of the Lepton thermal camera family, captured thermal images with higher resolution. Therefore, one pixel could cover about 2.3 x 2.3 mm\textsuperscript{2} area of the actual foot pointing out ulcer areas with a improved spatial accuracy. Since the camera module was attached to a small size breakout board, image capturing system (both software and firmware) could be easily modified depending on the requirement of the clinicians. However, more compact and user friendly device will improve the performance and accuracy of the data. The suggestions for that might include,
%
%

\begin{itemize}
\item Miniaturize the mechanical structure with enabling a focusing mechanism to the camera
\item Increase the stability of the structure
\item Take the video feed of feet from the thermal camera 
\end{itemize}

More than 80\% of high-temperature contralateral points which show the temperature differences more than or equal to 2.2\textsuperscript{0}C are found at the toe, metatarsal and heel areas. The algorithm used in the analysis process compares each and every contralateral pixel of the two feet and highlights the areas which had temperature difference greater than 2.2\textsuperscript{0}C. This gives an intuition regarding the location and the size of the high temperature area. Additionally, the validation step of the algorithm avoided the areas highlighted due to the low temperature of the other foot. Hence, this system could identify  high temperature wounds with 100\% accuracy. However, the algorithm can be further enhanced by introducing,

\begin{itemize}
\item More advanced machine learning concepts
\item Applying dynamic threshold values for the temperature comparison
\item Comparing the hotspots within the same foot
\end{itemize}

\section{CONCLUSION}
The overall system which was developed to profile the foot temperature which comprises of a hardware system and an analysis system has yielded results that concludes the effectiveness of NIR thermography and a robust image analysis modality which assists the clinicians to make decisions related to diabetic foot ulceration and other foot complications which result in temperature change.

This system can further be developed as discussed above which would accurately predict foot ulceration and monitor the progress of foot complications.

\addtolength{\textheight}{-12cm}   




\section*{ACKNOWLEDGMENT}
We extend our gratitude to all the academic staff of Department of Electronic and Telecommunication Engineering, University of Moratuwa for veteran assistance and guidance given in developing this system. 

We acknowledge the National Hospital of Sri Lanka for accommodating the testing and verification process by granting patient access.


\bibliographystyle{plain}
\bibliography{ref}
\nocite{*}

\end{document}